\documentclass[]{spie}  
\usepackage{graphicx}
\usepackage{amsmath}
\usepackage{cite}
\usepackage{bm}
\usepackage{gensymb}
\usepackage{graphicx}
\usepackage{multirow}
\usepackage{amsmath}
\usepackage{authblk}
\usepackage{adjustbox}
\usepackage{graphicx}
\usepackage{amssymb}
\usepackage{amsmath,amsfonts}
\usepackage{algorithm}
\usepackage{soul}
\usepackage[noend]{algpseudocode}
\usepackage{mathtools}
\usepackage[table]{xcolor}
\usepackage{booktabs}
\newcommand{\tabitem}{~~\llap{$\rightarrow$}~~}
\usepackage{caption}
\usepackage{subcaption}
\usepackage{diagbox}
\usepackage{hyperref}

\title{G-MIND: An End-to-End Multimodal Imaging-Genetics Framework for Biomarker Identification and Disease Classification}
\author[1]{Sayan Ghosal}
\author[2]{Qiang Chen}
\author[2,4]{Giulio Pergola}
\author[2]{Aaron L. Goldman}
\author[2]{William Ulrich}
\author[3]{Karen F. Berman}
\author[4,5]{Giuseppe Blasi}
\author[4]{Leonardo Fazio}
\author[4]{Antonio Rampino}
\author[4]{Alessandro Bertolino}
\author[2]{Daniel R. Weinberger}
\author[2]{Venkata S. Mattay}
\author[1]{Archana Venkataraman}
\affil[1]{Department of Electrical and Computer Engineering, Johns Hopkins University, USA} 
\affil[2]{Lieber Institute for Brain Development, USA}
\affil[3]{Clinical and Translational Neuroscience Branch, NIMH, NIH, USA}
\affil[4]{Department of Basic Medical Sciences, Neuroscience and Sense Organs, University of Bari Aldo Moro, Italy}
\affil[5]{Azienda Ospedaliero-Universitaria Consorziale Policlinico, Bari, Italy}

\begin{document} 
  \maketitle 

\begin{abstract}
We propose a novel deep neural network architecture to integrate imaging and genetics data, as guided by diagnosis, that provides interpretable biomarkers. Our model consists of an encoder, a decoder and a classifier. The encoder learns a non-linear subspace shared between the input data modalities. The classifier and the decoder act as regularizers to ensure that the low-dimensional encoding captures predictive differences between patients and controls. We use a learnable dropout layer to extract interpretable biomarkers from the data, and our unique training strategy can easily accommodate missing data modalities across subjects. We have evaluated our model on a population study of schizophrenia that includes two functional MRI (fMRI) paradigms and Single Nucleotide Polymorphism (SNP) data. Using 10-fold cross validation, we demonstrate that our model achieves better classification accuracy than baseline methods, and that this performance generalizes to a second dataset collected at a different site. In an exploratory analysis we further show that the biomarkers identified by our model are closely associated with the well-documented deficits in schizophrenia.
\end{abstract}


\section{Introduction}
Neuropsychiatric disorders, such as autism and schizophrenia, are typically characterized by cognitive and behavioral deficits\cite{Zihl1998}. At the same time, these diseases also show high genetic heritability \cite{Lee2005}, which suggests an important link between genotypic variations and the observed phenotypic traits \cite{Erk2017}. Understanding this relationship might lead to targeted biomarkers and eventually better therapeutics. Non-invasive techniques like functional MRI (fMRI) and Single Neucleotide Polymorphism (SNP) are commonly used data modalities to capture the brain activity and genetic variations, respectively. However, integrating them in a single framework is hard due to their inherent complexity, high data dimensionality, and our limited knowledge about the underlying relationships.

Imaging-genetics has become an increasingly popular field of study to link these modalities. Data driven methods can be grouped into three categories. The first category uses multivariate regularized regression to model the effect of genetic variations on the brain activity \cite{Wang2012, Liu2014}. These methods rely on sparsity to identify an interpretable set of biomarkers; however, they do not incorporate clinical diagnosis, meaning that the biomarkers may not align with predictive group differences. The second category uses correlation analysis to identify associations between genetic variations and quantitative traits \cite{Chi2013, Meda2012, Pergola2017}. However, the representations are rarely guided by the clinical factors, and it is not clear how they can be extended to accommodate more than two data modalities.
Finally, the recent works of \cite{Batmanghelich2016, Ghosal2019} use probabilistic modelling and dictionary learning, respectively, to integrate imaging, genetics, and diagnosis. The generative nature of these methods makes it harder to integrate additional data modalities. However, the field is moving towards multimodal imaging acquisitions to capture different snapshots of the brain all of which may have link to the genotype. Another limitation is that none of the above methods can handle the problem of missing data. With the growing emphasis on big datasets comes the challenge of missing data modalities. Traditionally, missing data has been managed by removing subjects from the analysis \cite{Kang2013}, which does not make use of all the information. In this paper we introduce a novel model and training strategy that accommodates these missing modalities to maximize the size and utility of the dataset.

The above limitations have motivated our use of deep learning, and specifically, the autoencoder architecture. First, the autoencoder provides a natural way to integrate new data modalities \cite{BenSaid2017} simply by adding new encoder-decoder branches. Mathematically, a new branch will introduce another term to the loss function but does not alter the optimization procedure (e.g., backpropagating gradients)\cite{Kingma2015AdamAM}. Second, missing data can easily be handled by freezing the affected part of the network \cite{Jaques2018} and updating the remaining weights. This simplicity is in stark contrast to the classical methods, where the entire model and optimization procedure must be changed for each new modality and missing data configuration. Third, the latent encoding provides a data-driven feature space that can be used for patient/control classification. Again, this is in contrast to classical approaches, which are highly dependent on hand-crafted feature. Finally, the classifier part of our model guides the autoencoder to extract clinically interpretable features that are representative of the disease.  

In this paper we introduce the \textbf{G}enetic and \textbf{M}ultimodal \textbf{I}maging data using \textbf{N}eural-network \textbf{D}esigns (\textbf{G-MIND}) framework to identify predictive biomarkers from neuroimaging and genetics data for disease diagnosis. We use a coupled autoencoder and classifier to learn a shared latent space among all the input modalities that is representative of the population differences between patients and controls. We also incorporate a learnable dropout layer \cite{Gal} by which the model selects a random subset of input features to pass to the encoder. The feature importances are captured in the probability of dropout, which is learned via backpropagation. We evaluate G-MIND on a study of schizophrenia that includes two task fMRI paradigms and SNP data. Our method achieves better classification performance than standard baselines and also identifies clinically relevant biomarkers. We further applied G-MIND to a cross-site dataset without any fine tuning to show the transferability of our model.

\section{The Multi-modal Encoder-Decoder Framework}
\begin{figure}[!t]
\fbox{
    \centering
    \includegraphics[width=0.9\columnwidth,keepaspectratio]{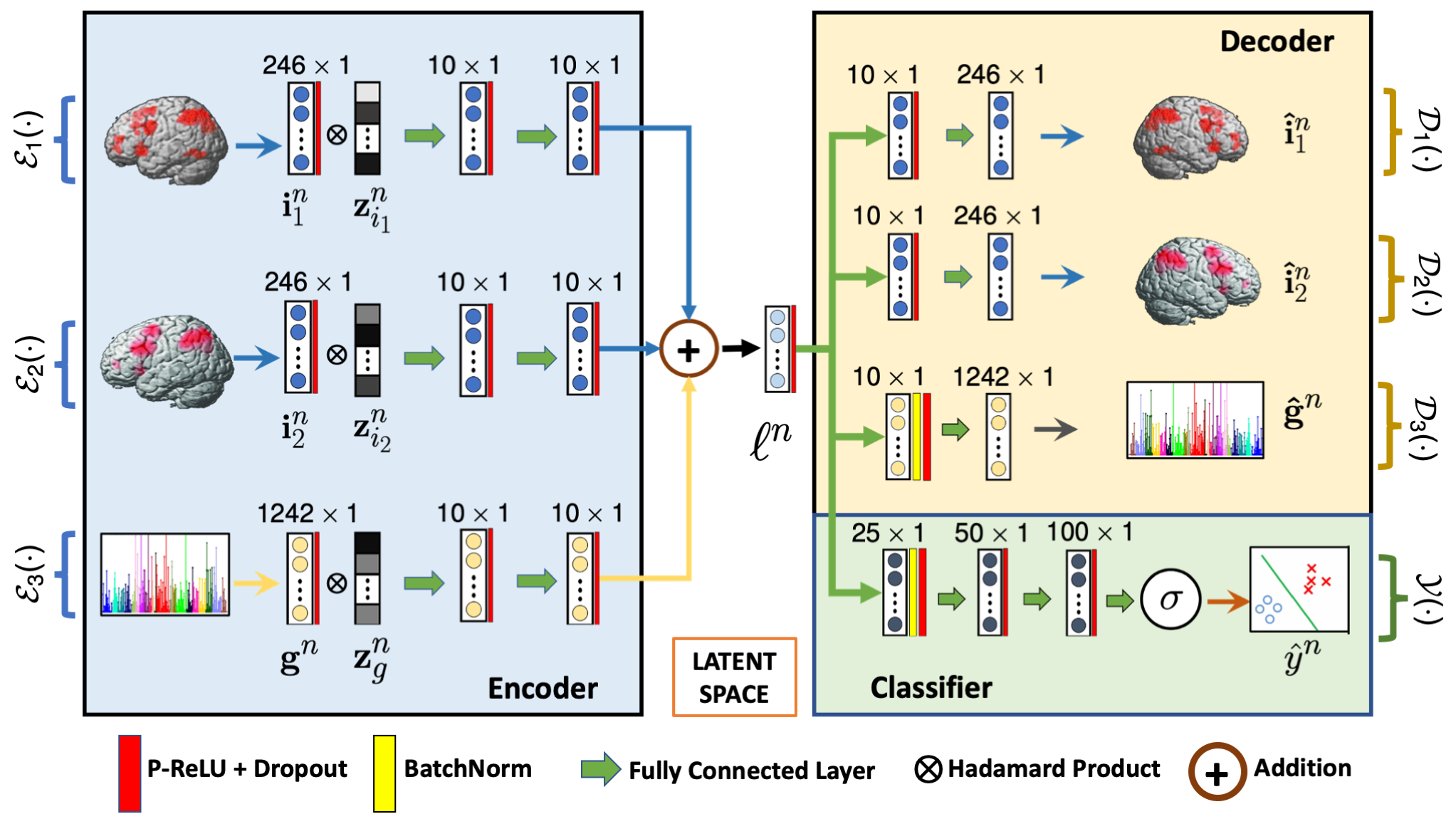}}
    \caption{G-MIND architecture. The inputs $\{\mathbf{i}_1, \mathbf{i}_2\}$ and $\{\mathbf{g}\}$ corresponds to the two imaging modalities and genetic data, respectively. $\mathcal{E}_i(\cdot)$ and $\mathcal{D}_i(\cdot)$ captures the encoding and decoding operations, and $\mathcal{Y}(\cdot)$ captures the classification operation. $\mathbf{z}_{i}$  is the learnable dropout mask, and $\mathbf{\ell}^n$ is the low dimensional latent space.}
    \label{main}
\end{figure}
Fig. \ref{main} illustrates our full model. The inputs $\mathbf{i}_1^n$ and $\mathbf{i}_2^n$ denotes the input imaging modalities for subject~$n$. In our case $\mathbf{i}_1^n$ and $\mathbf{i}_2^n$ are activation maps from two different fMRI paradigms. The input $\mathbf{g}^n$ represents the SNP genotype, and $y^n$ is a binary class label (patient or control). Let $N_1$, $N_2$, and $N_g$ denote the number of subjects from whom we have the corresponding imaging or genetic modality. Let $R$ be the total number of ROIs in brain, and $G$ be the total number of SNPs. The imaging data has the dimensionality $\mathbf{i}_1^n, \mathbf{i}_2^n \in \mathbf{R}^{R\times1}$, and the genetic data has the dimensionality $\mathbf{g}^n \in \mathbf{R}^{G\times1} $. We jointly model the imaging and genetic modalities using a auto-encoder framework. The first layer of the encoder incorporates a learnable dropout parameterized by $\mathbf{p}_m$ for each modality~$m$. We use the resulting low dimensional representation $\ell^n$ for subject classification.
\subsection{\textbf{Feature Importance using Learnable Dropout}}
The standard Bernoulli dropout independently drops nodes using a fixed probability. The Bayesian interpretation of dropout has a close resemblance with Bayesian feature selection \cite{gal2016dropout}, however, the user must fix the dropout probability \textit{a priori}. Here we wish to learn these vales, so we reparameterize the Bernoulli dropout mask with a Gumbell-Softmax \cite{Jang2016} distribution. This continuous relaxation of Bernoulli random variable enables us to update the dropout probabilities while training the network. During each forward pass through the network we sample random variables $\mathbf{z}_{i_1}^n,\mathbf{z}_{i_2}^n \in \mathbf{R}^{R \times 1}$, and $\mathbf{z}_g \in \mathbf{R}^{G\times 1}$ for imaging and genetic data, respectively, from a Gumbell-Softmax distribution and use it as a dropout mask for patient~$n$:
\begin{align}\label{repara}
    \mathbf{z}_{i_1}^n =\sigma \left(\frac{\log(\mathbf{p}_{i_1}) - \log(1 - \mathbf{p}_{i_1}) + \log(\mathbf{u}_{i_1}^n) - \log(1-\mathbf{u}_{i_1}^n)}{t}\right)
\end{align}
where $\mathbf{u}_{i_1}^n$ is a random vector sampled from $Uniform(0,1)$, the parameter $t$ (temperature) controls the extent of relaxation from the Bernoulli distribution and $\mathbf{p}_m$ captures the probabilities with which the features of modality~$m$ are selected. As seen in Eq. (\ref{repara}) when the probability $\mathbf{p}_{m_k}$ is close to $1$ that feature will be selected most of the time, as compared to a feature whose probability is close to $0$. We further incorporate a sparsity penalty over the probabilities $p_{mk}$ via the KL divergence $KL\left(Ber(q)||Ber(p_{mk})\right)$ where $q$ is a hyperparameter fixed to $0.001$. Effectively, this term encourages $p_{mk}$ towards zero for all features.
\subsection{\textbf{Multimodal Latent Encoding}}
The encoder learns a nonlinear latent space that is shared between all the data modalities. As shown in Fig.~\ref{main} we encode the data following the dropout using a cascade of fully connected layers followed by a PRelu activation \cite{He}. Unlike standard autoencoder based networks we couple the low dimensional representations of each data modality to leverage the common structures shared between them. The latent embedding $\ell^n$ is computed as
\begin{align}\label{enc}
    \mathbf{\ell}^n = \frac{1}{M_n}\left(\mathcal{E}_1(\mathbf{i}_1^n,\mathbf{z}_{i_1}^n) + \mathcal{E}_2(\mathbf{i}_2^n,\mathbf{z}_{i_2}^n)+\mathcal{E}_g(\mathbf{g}^n,\mathbf{z}_g^n)\right)
\end{align}
 Here $\mathcal{E}_i(\cdot)$ represents the encoding operation for modality~$m$, and $M_n$ is the number of modalities present for subject~$n$. As seen in Eq. (\ref{enc}), our latent representation is the sum of the individual projections, scaled by the amount of available data~$M_n$. This fusion strategy encourages the latent encoding for an individual patient to have a consistent scale, even when constructed using a subset of the modalities.
\subsection{\textbf{Data Reconstruction}}
The decoder reconstructs the data from the latent representation to ensure that the encoder is preserving sufficient information about the inputs. We use fully connected layers along with PRelu, dropouts, and batchnorm for decoding. Mathematically, the autoencoder loss is the $l_2$ norm between the input and reconstruction:
\begin{align*}
    \sum_{n=1}^{n_1}||\mathbf{i}^N_1 - \mathcal{D}_1(\ell^n) ||_2^2 + \sum_{n=1}^{N_2}||\mathbf{i}^n_2 - \mathcal{D}_2(\ell^n) ||_2^2 +
    \sum_{n=1}^{N_g}||\mathbf{g}^n - \mathcal{D}_3(\ell^n) ||_2^2
\end{align*}
where $\mathcal{D}_m(\cdot)$ is the decoding operation for modality~$m$.
\subsection{\textbf{Disease Classification}}
The final piece of our network is a classifier for disease prediction, which will encourage the dropout mask and latent embeddings to select discriminative features from the data. We employ fully connected layers, and a cross entropy loss for classification: $-\sum_{n=1}^{N} \left(y^n\log(\hat{y}^n) + (1-y^n)\log(1-\hat{y}^n)\right)$, where $y$ is the original class label and $\hat{y}^n$ is the predicted class label.

Our combined G-MIND objective function can be written as follows:
\begin{align}\label{full}
    \mathcal{L}(\mathbf{i}_1,&\mathbf{i}_2,\mathbf{g}) =     \lambda_1\sum_{n=1}^{N_1}||\mathbf{i}^n_1 - \mathcal{D}_1(\ell^n) ||_2^2 + \lambda_2\sum_{n=1}^{N_2}||\mathbf{i}^n_2 - \mathcal{D}_2(\ell^n) ||_2^2 \nonumber \\ 
    &+\lambda_3\sum_{n=1}^{N_g}||\mathbf{g}^n - \mathcal{D}_3(\ell^n) ||_2^2 - \lambda_4\sum_{n=1}^{N}\left(y^n\log(\hat{y}^n) + (1-y^n)\log(1-\hat{y}^n)\right) \nonumber \\
    &+ \lambda_5\sum_{m=1}^3\sum_{k}KL(Ber(q)||Ber(p_{mk})
\end{align}
where $N$ is the total number of subjects. The parameters $\{\lambda_1, \lambda_2, \lambda_3 \}$ control the contributions of the data reconstruction error, $\lambda_4$ controls the contribution of classification error, and $\lambda_5$ regularizes the sparsity on $\mathbf{p}_m$.

\begin{figure*}[!t]
 \begin{minipage}[c]{0.5\columnwidth}
 \fbox{
 \centering
    \includegraphics[width=\columnwidth, keepaspectratio]{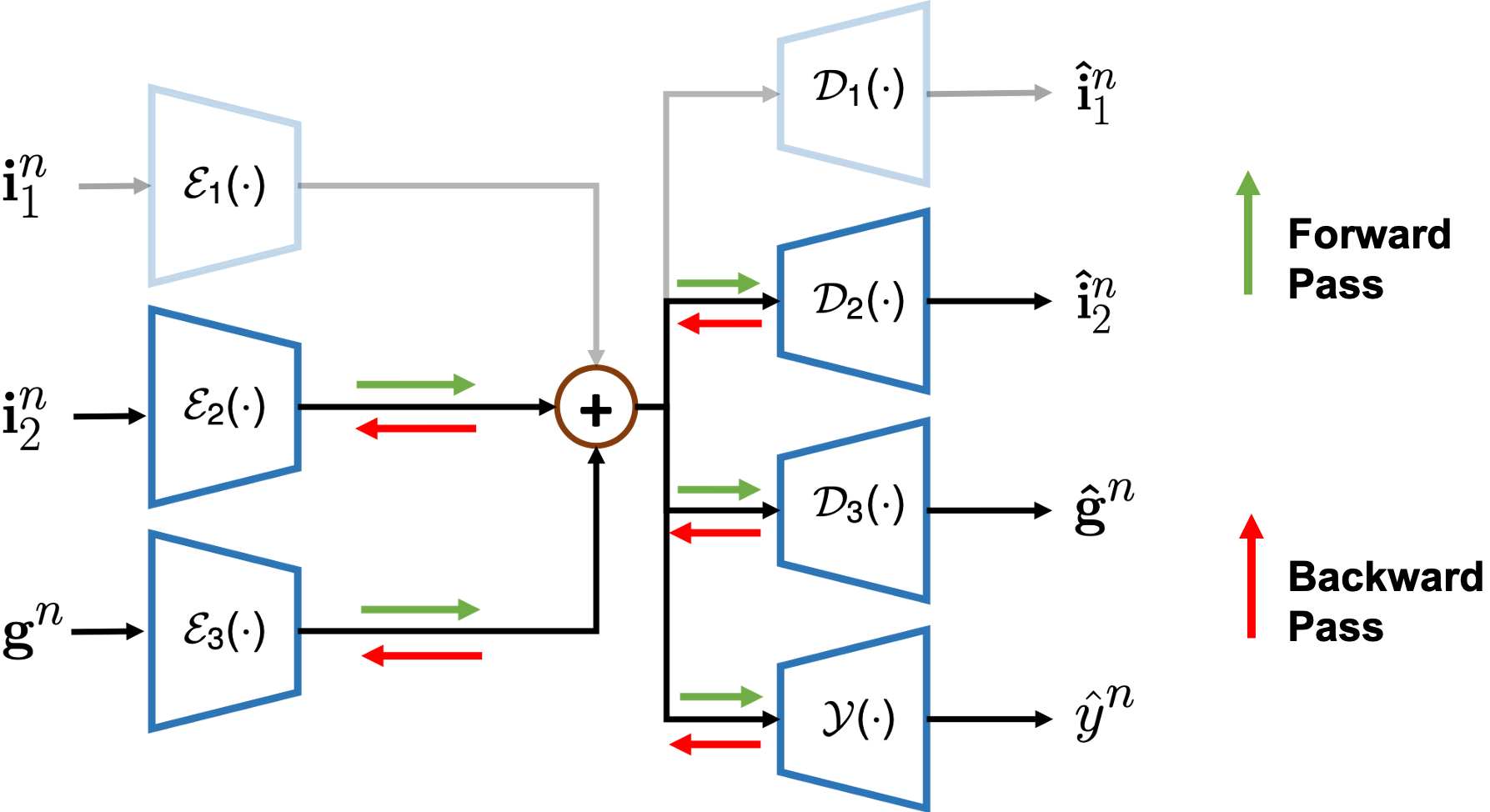}}
       
\end{minipage}
~
 \begin{minipage}[c]{0.4\columnwidth}
\centering
\resizebox{\columnwidth}{!}{%
\renewcommand{\arraystretch}{1.2}
      \begin{tabular}{| c | c | c | c |}
    \hline
    \multirow{2}{*}{ Institution } &
      \multicolumn{3}{c|}{ Modalities }\\
      \cline{2-4}
    & \ \ NBack \ \ & \ \ SDMT \ \  & \ \ SNP \ \ \\[1pt]
    \hline
    \ \ LIBD \ \ & 160 & 110 & 210\\
    \hline
    BARI  & 97 & \cellcolor{gray!100} & 97\\
    \hline
  \end{tabular}
  }
 \end{minipage}
     \begin{minipage}[t]{.5\columnwidth}
    \caption{Information flow during the forward pass (green) and backward pass (red) when $\mathbf{i}_1^n$ is absent.}
    \label{miss}
    \end{minipage}%
    \hfill%
    \begin{minipage}[t]{.4\columnwidth}
    \captionof{table}[foo]{{The number of subjects present for each modality from the two institutions. Note that the SDMT task was not acquired for BARI}}
\label{subjects}
    \end{minipage}%
\end{figure*}

The summation in Eq. (\ref{full}) enables G-MIND to handle missing data. For example if $\mathbf{i}_1^n$ is not available for subject $n$, then the gradients with respect to encoder $\mathcal{E}_1(\cdot)$ and decoder $\mathcal{D}_1(\cdot)$ will be zero. As illustrated in Fig. (\ref{miss}), information will flow into and out of the latent space through the other network branches, and will only be used to update those parameters.
\subsection{\textbf{Prediction on New Data}} 
During training, we learn the encoder, decoder, and classifier weights, along with the probabilistic masks~$\mathbf{p}_m$ by minimizing Eq.~(\ref{full}). We then threshold the probabilistic mask $\mathbf{\hat{p}}_m = \left(\mathbf{p}_m > \tau_m\right) $ to select the most important features for reconstruction and classification. When testing on a new subject data, we premultiply the available modalities by the thresholded dropout mask, i.e., $\hat{\mathbf{i}}_1^n = \mathbf{i}_1^n \otimes \mathbf{\hat{p}}_{i_1} $.
The masked input $\hat{\mathbf{i}}_1^n$ is sent through encoder and the classifier for diagnosis. We do not use the learned dropout procedure during testing, since different samples of $\mathbf{z}_m^n$ may lead to a different diagnosis, whereas our goal to obtain a deterministic label for each subject.
\subsection{\textbf{Implementation Details}}
We set the regularization parameters of our model $\{\lambda_1, \lambda_2, \lambda_3, \lambda_4, \lambda_5\}$ as $10^{-\beta_i}$ where $\beta_i$ is selected such that $\lambda_i$ multiplied by the appropriate loss term lies within the same order of magnitude (1---10). This criterion is intuitive (i.e., equal importance is given to both the imaging and genetic data), and it is not performance driven (i.e., we do not cherry-pick the values to optimize prediction accuracy). The corresponding values for all the experiments are:  $\lambda_1 = 0.1, \lambda_2 = 0.1, \lambda_3=0.01, \lambda_4=0.1,$ and $\lambda_5=0.01$. We fix the Bernoulli probability, to $q=0.001$ and the temperature variable to $t=0.1$. Based on 10-fold cross validation results we fix all detection threshold values to $\tau_{i}=0.1$. The architecture of our model (layer sizes and nonlinearities) is shown in Fig.~\ref{main}. 
\subsection{\textbf{Baseline Comparison Methods}}
We compare G-MIND to classical machine learning techniques and architectural variants that omit key features.
\begin{itemize}
    \item \textbf{Multimodal Support Vector Machine (SVM):} We construct a linear SVM classifier after concatenating all the data modalities $[\mathbf{i}_1^T,\mathbf{i}_2^T,\mathbf{g}^T]^T$. 
    Notice that this model cannot handle missing data. Therefore, we fit a multivariate regression to impute missing imaging modalities based on the available one for each subject. For example if $\mathbf{i}_1^n$ is absent, we impute it as: $\mathbf{i}_1^n = \boldsymbol{\beta}^*\mathbf{i}_2^n$, where $\boldsymbol{\beta}$ is the regression coefficient matrix obtained from training data.
    We use a grid search method to find the best set of hyper-parameters. Notice that this tuning provides an \textit{added advantage} for SVM over G-MIND.
    \item \textbf{Multimodal CCA + RF:}
    Canonical correlation analysis (CCA) identifies bi-multivariate associations between imaging and genetics data. This approach is similar to our coupled latent projection, but the traditional CCA does not accommodate more than two data modalities. In order to overcome this we concatenate the imaging features obtained from two experimental paradigms and perform CCA with the genetics data. We then construct a random forest classifier based on the latent projections. We use the same approach for data imputation and to find the best set of hyperparameters.
    \item \textbf{Encoder Only:}
    We compare our model to an ANN architecture based on the encoder and the classifier of G-MIND. This comparison will show us importance of using the decoder and the learnable dropout layer.
    \item \textbf{Encoder+Dropout:}
    We compare our model to another ANN architecture where we only used the encoder, the classifier, and the learnable dropout layer. This experiment will show us the performance improvement from including a decoder.
    Based on our $10$-fold cross validation we fix the learned dropout threshold values to $\{\tau_{i_1}=0.05, \tau_{i_2} = 0.05, \tau_g = 0.1\}$. 
    
\end{itemize}
\section{Experimental Results}
\subsection{\textbf{Data and Preprocessing}}
Our first dataset includes two task fMRI paradigms and SNP data provided by Lieber Institute for Brain Development (LIBD) in Baltimore, MD, USA. The first fMRI paradigm is a working memory task (Nback) that alternates between 0-back and 2-back trial blocks. During the 0-Back task the subjects is asked to press a number shown on the screen, and during the 2-back task the subjects are instructed to press the number shown 2 stimuli previously.  The second fMRI paradigm is an event-based simple declarative memory task (SDMT) which involves incidental encoding of complex visual scenes. Our replication dataset includes just Nback and SNP data acquired at the University of Bari Aldo Moro, Italy (BARI). The distribution of the subjects is shown in Table~\ref{subjects}. All fMRI data was acquired on 3-T General Electric Sigma scanner (EPI, TR/TE $=$ 2000/28 msec; flip angle $=$ 90; field of view $=$ 24 cm, res: 3.75 mm in x and y dimensions and 6 mm in the z dimension for NBack and 5 mm for SDMT). FMRI preprocessing include slice timing correction, realignment, spatial normalization to an MNI template, smoothing and motion regression. SPM12 is used to generate activation and contrast maps for each paradigm. We use the Brainnetome atlas \cite{Fan2016} to define 246 cortical and subcortical regions. The input to our model is the contrast map over these ROIs. 

In parallel, genotyping was done using variate Illumina Bead Chips including 510K/ 610K/660K/2.5M. Quality control and imputation were performed using PLINK and IMPUTE2 respectively. The resulting 102K linkage disequilibrium independent SNPs are used to calculate the polygenic risk score of schizophrenia via a log-odds ratio of the imputation probability for the reference allele \cite{Chen2018}. By selecting $P<10^{-4}$, we obtain $1242$ linkage disequilibrium independent SNPs. As a preprocessing step, we remove the effect of age, IQ, and education from the imaging modalities and we have mean centred all the data modalities.
\begin{figure*}[!t]
 \begin{minipage}[c]{0.45\columnwidth}
\centering
\resizebox{\columnwidth}{!}{%
      \renewcommand{\arraystretch}{1.5}
  \renewcommand{\arraystretch}{1.5}
  \begin{tabular}{|c|c|c|c|c|c|}
    \hline
          \begin{tabular}[c]{@{}c@{}} \backslashbox{\textbf{  Method }}{\textbf{Perf }}\end{tabular}  &
      \begin{tabular}[c]{@{}c@{}}\ \  Sens \ \ \end{tabular}  &
      \begin{tabular}[c]{@{}c@{}}\ \  Spec \ \ \end{tabular} &
      \begin{tabular}[c]{@{}c@{}}\ \  Acc \ \ \end{tabular} &
      \begin{tabular}[c]{@{}c@{}}\ \  Auc \ \ \end{tabular}\\
          \hline
    \begin{tabular}[c]{@{}c@{}}\  SVM \  \end{tabular} & 0.66 & 0.47 & 0.58 & 0.55\\
    \hline
    \begin{tabular}[c]{@{}c@{}}\ CCA+RF \  \end{tabular} & 0.15 & \textbf{0.92} & 0.51 & 0.56\\
    \hline
    \begin{tabular}[c]{@{}c@{}}\ Encoder Only \ \end{tabular} & 0.57 & 0.57 & 0.57 & 0.59\\
    \hline
    \begin{tabular}[c]{@{}c@{}}\  Encoder + Dropout \ \end{tabular}
    & 0.61 & 0.56 & 0.59 & 0.62
    \\
    \hline
    \begin{tabular}[c]{@{}c@{}}\ G-MIND  \ \end{tabular}
    & \textbf{0.75} & 0.58 & \textbf{0.67} & \textbf{0.68}
    \\
    \hline
  \end{tabular}
  }
\end{minipage}
~
 \begin{minipage}[c]{0.5\columnwidth}
\centering
    \includegraphics[width=\columnwidth, keepaspectratio]{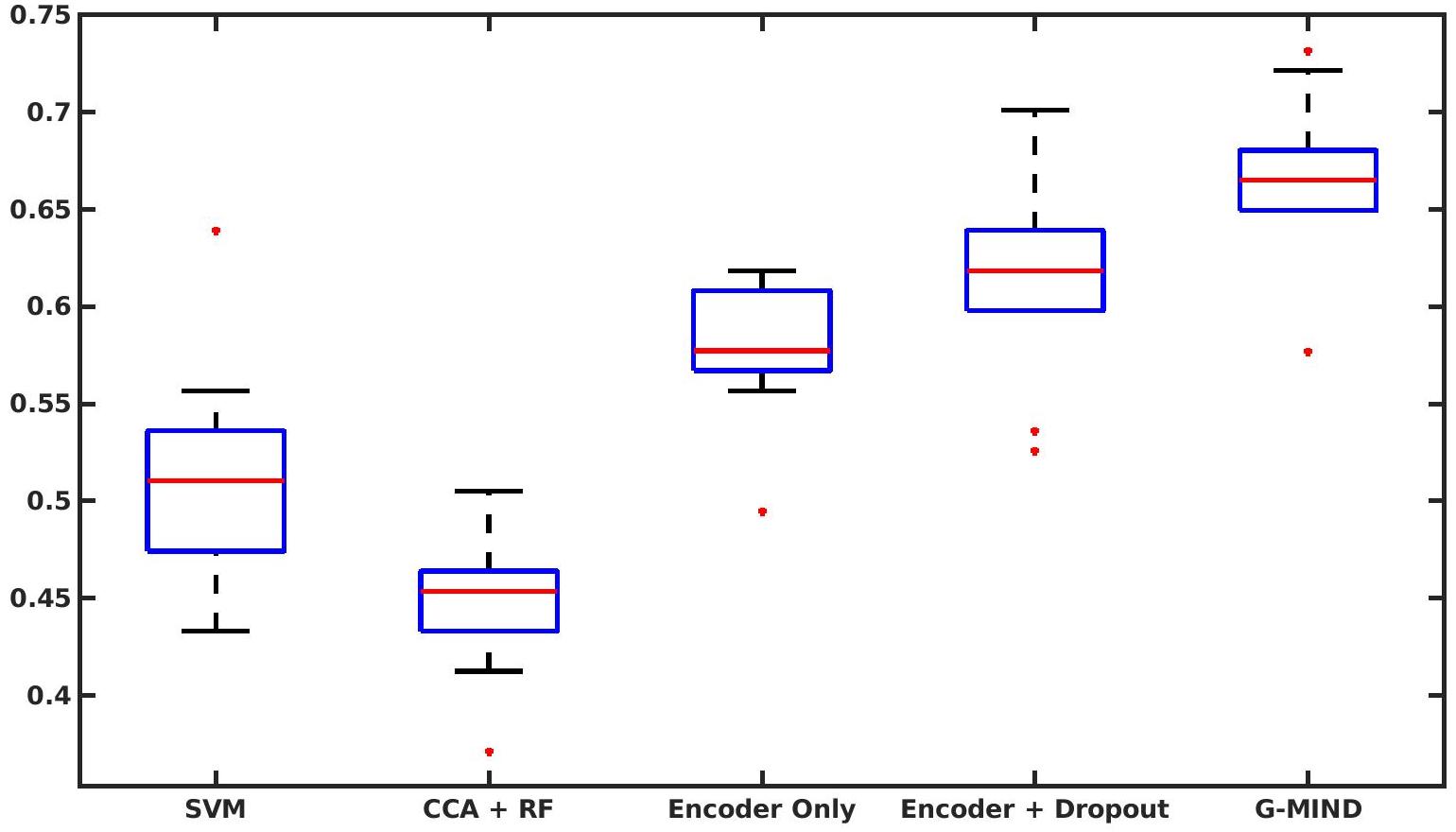}
 \end{minipage}
     \begin{minipage}[t]{.4\columnwidth}
    \captionof{table}[foo]{{Testing performance of each method on LIBD during 10 fold cross validation.}}
    \label{t1}
    \end{minipage}%
    \hfill%
    \begin{minipage}[t]{.5\columnwidth}
    \caption{Distribution of accuracies by the models trained in all 10 CV folds, when directly evaluated on BARI.}
    \label{box}
    \end{minipage}%
\end{figure*}
\subsection{{Model Performance}}Table \ref{t1} quantifies the 10-fold testing performance of all the methods on multimodal data obtained from LIBD. We can clearly see that G-MIND achieves the best overall accuracy. Even in the presence of missing data our multi modal approach can successfully extract meaningful information from all the data modalities that are essential for diagnosis prediction. Our results also show the importance of the decoder and the dropout layer.

In order to show the generalizability of our method we trained our model on LIBD data and tested it
without fine tuning on a cross site dataset from BARI. This experiment
captures the transference property of our model. We note that the SDMT task was not acquired at BARI, so the corresponding branch of G-MIND is not used. We evaluate the 10 best models obtained from the 10
different folds to run this experiment. Fig. \ref{box} shows the distribution of accuracies
of all the models in the form of a boxplot. Here we can see that our method shows
the best transference property compared to all the baselines. This is an interesting result as it shows the robustness of our model against data acquisition noise and population specific noise.
This performance gain further suggests that the learnable dropout mask can identify robust set of features
that are most predictive of the disease.

\begin{figure}[!b]
      \centering
      \fbox{ 
 \includegraphics[width=0.6\columnwidth, keepaspectratio]{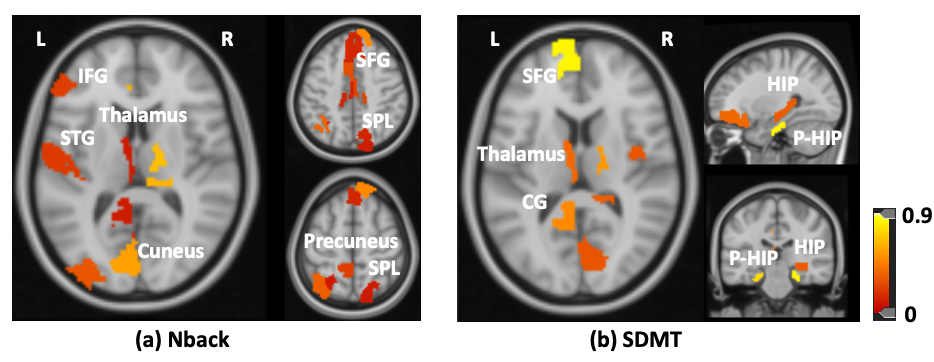}}
    \caption{The representative set of brain regions as captured by the dropout probabilities $\{\mathbf{p}_1, \mathbf{p}_2\}$. The color bar denotes the median value across 10 folds.}
    \label{im_bio}
\end{figure}
\begin{figure}[!t]
    \centering
    \fbox{ 
    \includegraphics[width=0.7\columnwidth, keepaspectratio]{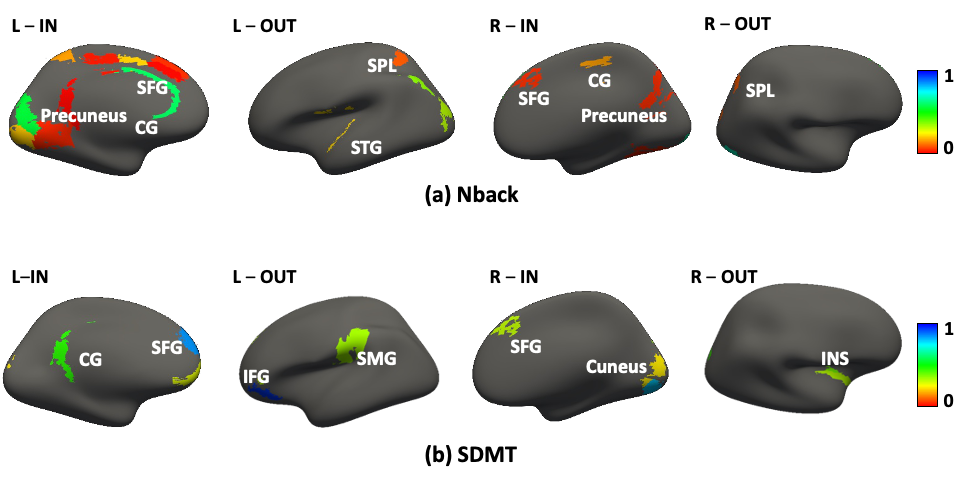}}
    \caption{The surface plot of the brain regions as captured by the dropout probabilities $\{\mathbf{p}_1, \mathbf{p}_2\}$. The color bar denotes the median value across 10 folds. From \textbf{Left} to \textbf{Right} the images are internal surface of left hemisphere (\textbf{L--IN}), external surface of left hemisphere (\textbf{L--OUT}), internal surface of right hemisphere (\textbf{R--IN}), and external surface of right hemisphere (\textbf{R--OUT}).}
    \label{im_surf}
\end{figure}
\begin{figure}[!b]
    \centering
    \fbox{ 
    \includegraphics[width=0.57\columnwidth, keepaspectratio]{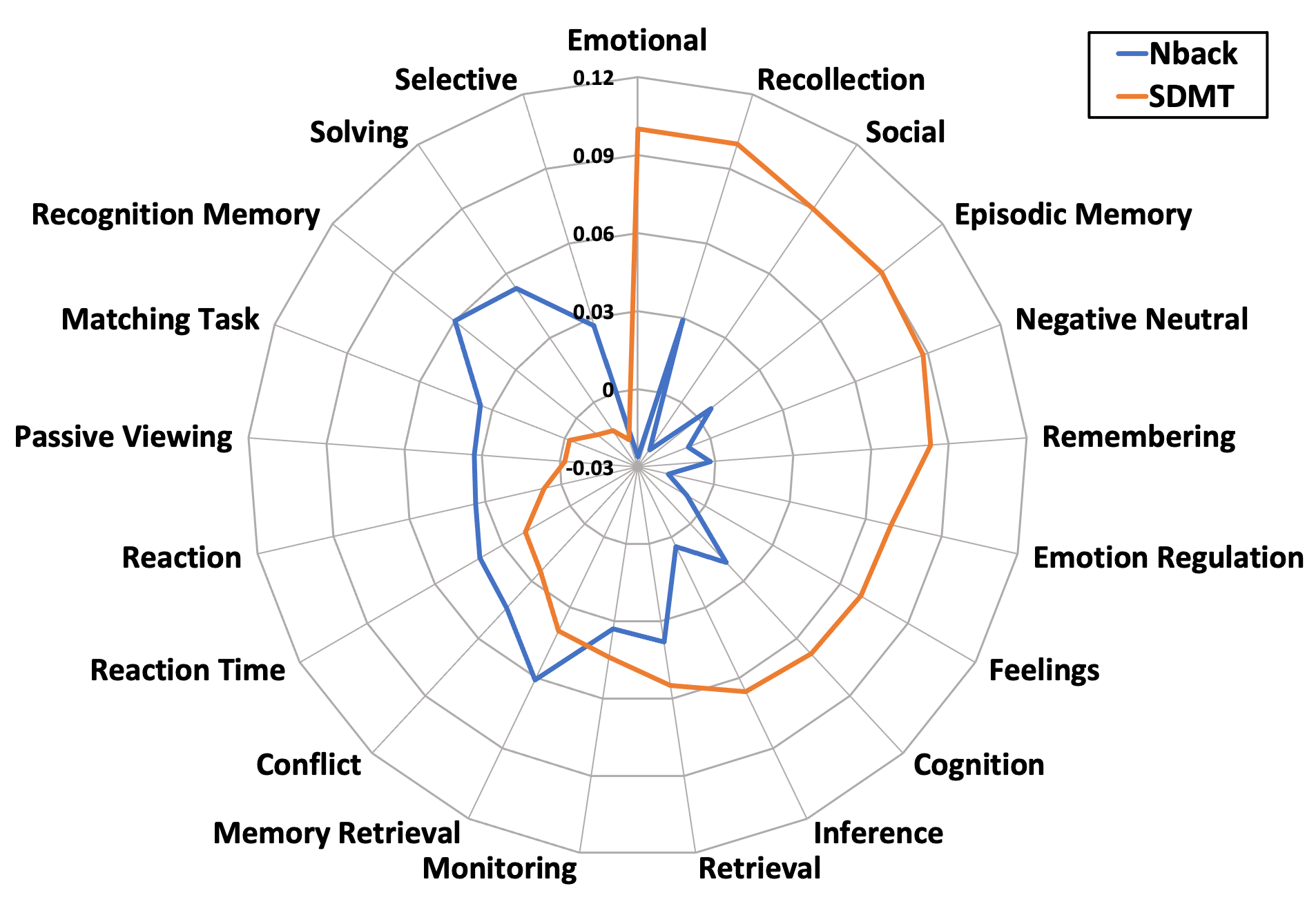}}
    \caption{The level of association with different cognitive states of all the brain regions identified by our model as found in the Neurosynth database.}
    \label{cog_st}
    \end{figure}
\subsection{Analysis of Imaging Biomarkers}
Fig. \ref{im_bio} illustrates the most important set of brain regions as identified by the median concrete dropout probability maps $\{\mathbf{p}_{i_1},  \mathbf{p}_{i_2}\}$ across the 10 validation folds. We further show a more global picture of the high importance brain regions as a surface plot in Fig. \ref{im_surf}. Both from, Fig. \ref{im_bio} and Fig. \ref{im_surf} for the Nback task we can see regions that include superior frontal gyrus (SFG),and inferior frontal gyrus (IFG), which are know to sub-serve  executive  cognition \cite{Callicott2003}. Moreover, we can see regions (SFG, IFG) from dorsolateral prefrontal cortex \cite{Callicott2003}  and regions (SPL, STG) from posterior parietal cortex that overlaps with the fronto-parietal network which is known to be altered in schizophrenia. Further clusters incorporate components of the default mode network also implicated in schizophrenia \cite{Sambataro2010}. The SDMT biomarkers implicate the hippocampal, parahippocampal, superior frontal regions along with the anteromedial thalamus which are also affected in schizophrenia \cite{Rasetti2014}. These regions control executive cognition and memory encoding and that are also known to be associated with the disorder.

We further use Neurosynth \cite{TorD.2011} to decode the higher order brain states of the the biomarkers associated with Nback and SDMT tasks. This analysis allows us to quantitatively compare the selected brain regions with previously published results and gives us a level of association with different brain states as identified by other studies. Fig. \ref{cog_st} shows the Neurosynth terms that are strongly correlated with our  biomarkers. We note that the terms associated with the Nback task corresponds to recognition and solving while the brain states for SDMT are associated with emotions and memory encoding.
These results provide further evidence that G-MIND can extract potential imaging biomarkers that are highly relevant to the task and the disorder under study.

\begin{figure*}[!t]
 \begin{minipage}[!t]{0.5\columnwidth}
 \centering
    \includegraphics[width=0.95\columnwidth, keepaspectratio]{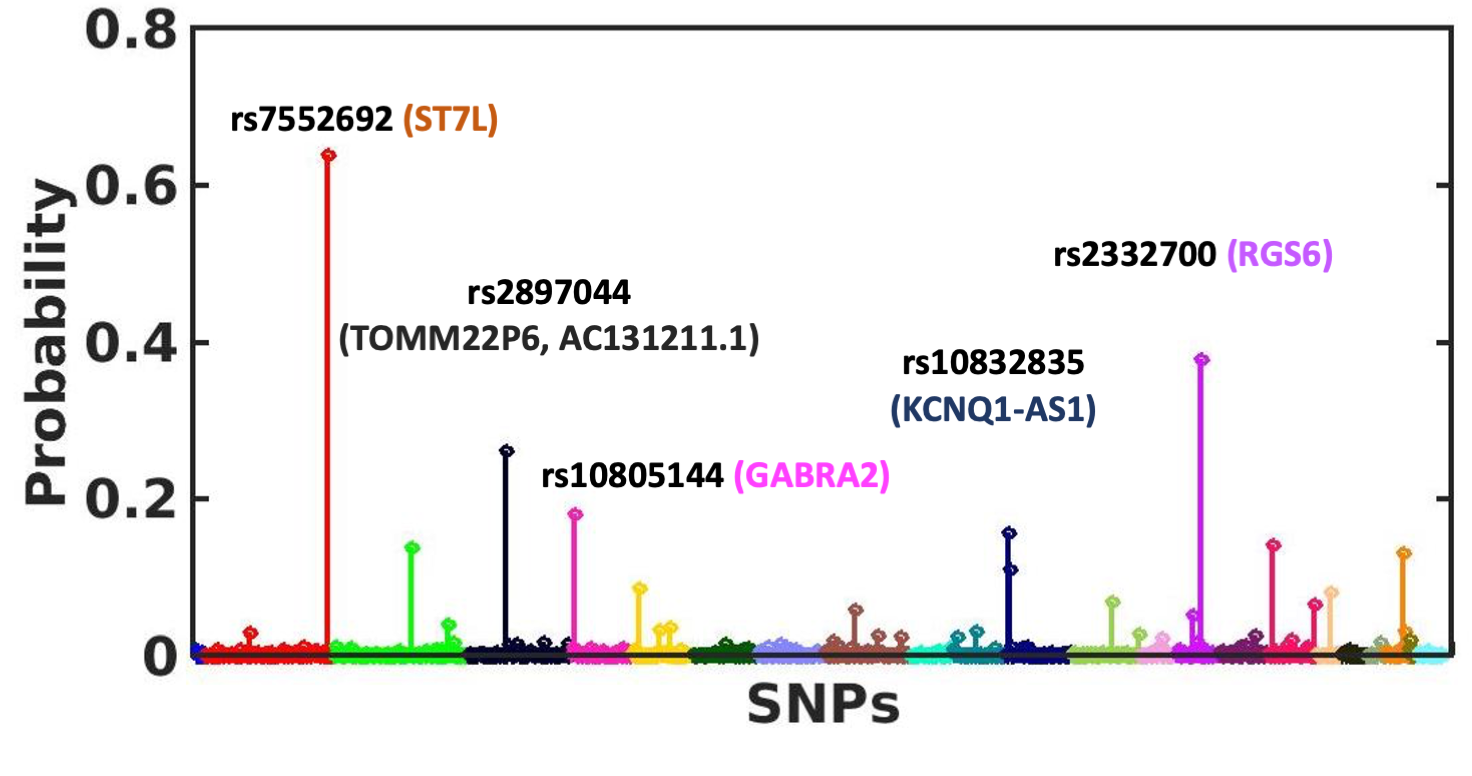}
\end{minipage}
~
 \begin{minipage}[!t]{0.5\columnwidth}
 \resizebox{\columnwidth}{!}{%
\centering
    \renewcommand{\arraystretch}{1.2}
 \begin{tabular}{ | l | c | } 
\hline
\  Biological Processes & \ \ FDR \ \ \\
\hline
\  Central nervous system development & 0.005\\ 
 \  \quad \tabitem Nervous system development & 0.001\\ 
 \  \quad \quad \tabitem System development.& 0.005\\ 
 \  Generation of neurons & 0.005\\ 
 \  \quad \tabitem Neurogenesis & 0.004
\\ 
 \ \begin{tabular}{l@{}}Regulation of calcium ion \\ transport into cytosol \end{tabular} & 0.04\\
 \ \tabitem Regulation of sequestering of calcium ion & 0.008\\
\hline
\end{tabular}
}
 \end{minipage}
 \\
     \begin{minipage}[t]{.47\columnwidth}
         \caption{The median importance map of all the SNP across and their overlapping genes across the 10 folds.}
         \label{snp}
    \end{minipage}%
    \hfill%
    \begin{minipage}[t]{.47\columnwidth}
    \captionof{table}[foo]{{The enriched biological processes and their level of significance obtained via GO enrichment analysis.}}
    \label{t2}
    \end{minipage}%
\end{figure*}
\begin{figure}[!b]
    \centering
    \fbox{ 
    \includegraphics[width=0.7\columnwidth, keepaspectratio]{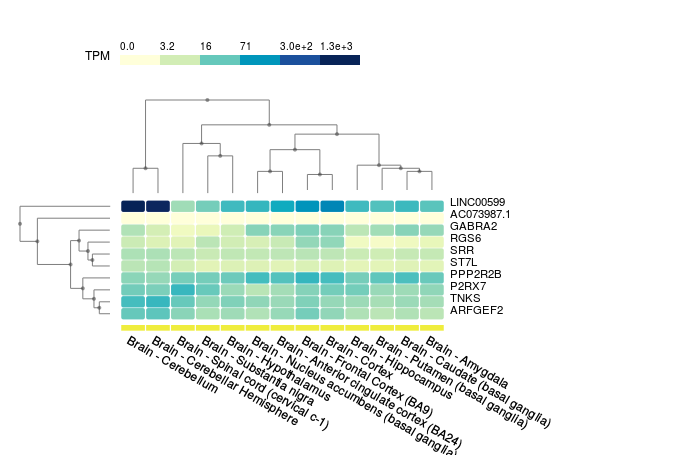}}
    \caption{The gene expression pattern of the selected set of genes in different brain tissues based on the GTEx database. Higher level of a gene expression in a brain tissue imply that alteration in that gene may have a stronger effect on those specific brain regions.}
    \label{gtex}
    \end{figure}
Fig. \ref{snp} shows the importance map across the $1242$ SNPs as computed by the median $\mathbf{p}_g$ across the 10 folds. We annotated each SNP based on its overlapping or nearest gene  as  found  from  the  SNP-nexus  web  interface \cite{DayemUllah2018}. In addition, we ran a gene ontology enrichment analysis of the overlapping genes of the top 300 SNPs to identify the enriched biological processes \cite{Mi2019}. This enrichment analysis gives us a way to identify the set of over-represented genes in a biological pathway that may have an association with the disease phenotype.
Table \ref{t2} captures the most significant biological processes implicated by the set of SNPs, which include the \textit{nervous system development} \cite{Dean2009}, and \textit{calcium ion regulation} \cite{Berridge2013} which are known to be strongly associated with schizophrenia. As parallel to Neurosynth analysis, we perform a gene expression based analysis \cite{Lonsdale2013} {over the 10 overlapping (or nearest gene if there is no overlap) genes} of the top SNPs identified from our analysis. Here we use the GTEx database to identify the set of brain tissues where these genes show high levels of expression. {This exploratory analysis may help us to understand the \textit{cis}-effects of the SNPs and how they alter the functionalities of genes expressed in different tissues of the brain}. Figure \ref{gtex} shows the gene expression pattern of each gene across different brain tissues. Here, \textit{LINC00599} shows high expression levels in brain and are also known to be associated with schizophrenia \cite{Goes2015} and neuroticism \cite{Luciano2018}. 
{These findings show that the model can be used to explore potential genetic biomarkers and their interactions in a multivariate framework.}
\section{Conclusion}
We have presented G-MIND, a novel deep network to integrate multimodal imaging and genetic data for targeted biomarker discovery and class prediction. Our unique use of learnable dropout with a classification module helps us to identify discriminative biomarkers of the disease. Our unique loss function enables us to handle missing modalities while mining all the available information in the dataset. We demonstrate our framework on fMRI and SNP data of schizophrenia patients and controls from two different sites.
The improved performance of G-MIND across all the experiments shows the capability this model to build a comprehensive view about the disorder based on the incomplete information obtained from different modalities. We note that, our framework can easily be applied to other imaging modalities, such as structural and diffusion MRI simply by adding autoencoder branches. In future work we will develop a hybrid extension of G-MIND in which we incorporate pathway specific information into the deep learning architecture for better understanding of the disease propagation.
\paragraph{Acknowledgements:}
This work was supported by NSF CRCNS 1822575, and the National Institute of Mental Health extramural research program.
\paragraph{Previous Submission}
This work has not been published to any other publication venue. \\
\bibliography{article320}          
\bibliographystyle{sp}   

\end{document}